\documentclass[reprint,
superscriptaddress,
nofootinbib,
aps,
pra,
showkeys
]{revtex4-2}
\pdfoutput=1

\usepackage{amsthm, mathtools, amssymb}
\usepackage{braket, physics} 
\usepackage{nicefrac}

\usepackage{orcidlink}
\graphicspath{ {./images/} }

\usepackage{graphicx, placeins, float}
\usepackage{booktabs}
\usepackage[caption=false]{subfig}
\usepackage[export]{adjustbox}
\usepackage{comment}

\usepackage{xcolor}

\usepackage{hyperref}
\usepackage[nameinlink,capitalize]{cleveref}
\hypersetup{
	colorlinks = true,
	linkbordercolor = {white},
	linkcolor={purple},
	citecolor={purple},
	urlcolor={blue}
}

\Crefname{appendix}{Supplementary Material}{Supplementary Materials}

\DeclareMathOperator*{\argmax}{arg\,max}

\theoremstyle{definition}

\makeatletter
\def\thm@space@setup{\thm@preskip=0pt
\thm@postskip=0pt}
\makeatother
\newtheorem*{theorem*}{Theorem}
\makeatletter
\def\thm@space@setup{\thm@preskip=0pt
\thm@postskip=0pt}
\makeatother

\begin{document}
\title{Divide et impera: hybrid multinomial classifiers from quantum binary models}
\author{Simone Roncallo\,\orcidlink{0000-0003-3506-9027}}
	\email[Simone Roncallo: ]{simone.roncallo01@ateneopv.it}
	\affiliation{Dipartimento di Fisica, Università degli Studi di Pavia, Via Agostino Bassi 6, I-27100, Pavia, Italy}
	\affiliation{INFN Sezione di Pavia, Via Agostino Bassi 6, I-27100, Pavia, Italy}
	
\author{Angela Rosy Morgillo\,\orcidlink{0009-0006-6142-0692}}
	\email[Angela Rosy Morgillo: ]{angelarosy.morgillo01@ateneopv.it}
	\affiliation{Dipartimento di Fisica, Università degli Studi di Pavia, Via Agostino Bassi 6, I-27100, Pavia, Italy}
	\affiliation{INFN Sezione di Pavia, Via Agostino Bassi 6, I-27100, Pavia, Italy}

\author{Seth Lloyd\,\orcidlink{0000-0003-0353-4529}}
	\email[Seth Lloyd: ]{slloyd@mit.edu}
	\affiliation{Massachusetts Institute of Technology, Cambridge, MA 02139, USA}
	
\author{Chiara Macchiavello\,\orcidlink{0000-0002-2955-8759}}
	\email[Chiara Macchiavello: ]{chiara.macchiavello@unipv.it}
	\affiliation{Dipartimento di Fisica, Università degli Studi di Pavia, Via Agostino Bassi 6, I-27100, Pavia, Italy}
	\affiliation{INFN Sezione di Pavia, Via Agostino Bassi 6, I-27100, Pavia, Italy}
	
\author{Lorenzo Maccone\,\orcidlink{0000-0002-6729-5312}}
	\email[Lorenzo Maccone: ]{lorenzo.maccone@unipv.it}
	\affiliation{Dipartimento di Fisica, Università degli Studi di Pavia, Via Agostino Bassi 6, I-27100, Pavia, Italy}
	\affiliation{INFN Sezione di Pavia, Via Agostino Bassi 6, I-27100, Pavia, Italy}
	
\begin{abstract}
	We investigate how to combine a collection of quantum binary models into a multinomial classifier. We employ a hybrid approach, adopting strategies like one-vs-one, one-vs-rest and a binary decision tree. We benchmark each method, by emphasizing their computational overhead and their impact on the quantum advantage. By comparison against a classical binary model (generalized using the same approach), we show that the decision tree represents a cost-effective solution, achieving similar accuracies to other methods with an overhead at most logarithmic in the total number of classes.
\end{abstract}
\keywords{Quantum classifier; Quantum optical neuron; Quantum neural networks;}
\maketitle

\section{INTRODUCTION}
Classification is a task in machine learning where, using the prior information on a set of labelled examples, one can assign a label to a new, i.e.~unseen, sample. Most problems involve more than two classes; for example, recognizing handwritten digits, penguin species or healthcare data requires distinguishing among multiple categories, namely to learn a multinomial distribution. Classical implementations of classifiers include fully-connected and convolutional neural networks, as well as kernel methods~\citep{book:Goodfellow}.

Quantum machine learning replaces classical data processing with quantum operations, with the purpose to provide a computational advantage~\citep{lloyd2013quantum,cai2015entanglement,tacchino2019artificial,benatti2019continuous,mangini2020quantum,steinbrecher2019quantum,killoran2019continuous,sui2020review,zhang2021quantum,spall2022hybrid,stanev2023deterministic,wood2024kerr,spall2025training,hoch2025quantum,slabbert2025classical, sun2025scalable,sakurai2025quantum}. Many quantum classifiers are inherently binary, namely the classification score is derived from a single quantity, usually a probability. Examples include, variational quantum classifiers~\citep{schuld2020circuit,havlivcek2019supervised,maheshwari2021variational}, support vector machines~\citep{rebentrost2014quantum}, kernel methods~\citep{schuld2019quantum,hubregtsen2022training,bowie2023quantum}, pretty good measurements~\citep{sergioli2019new}, and SWAP-test classifiers~\citep{blank2020quantum,nagies2026enhancing}. Extending such architectural constraints to more than two classes without compromising the quantum advantage is not straightforward. Typically, sampling from a multinomial distribution of $K$ classes introduces a computational overhead of at least $\mathcal{O}(K)$ operations, i.e.~the same cost of an output layer of $K$ neurons in a classical neural network. When such modifications are not available, nor feasible, an alternative is to rely on post-processing techniques, e.g.~by decomposing the multiclass problem into a collection of binary tasks, each handled by an independent classifier~\citep{book:Bishop}.

In this paper, we address this problem in the context of the quantum optical classifier introduced in~\citep{roncallo2025quantum,roncallo2025shallow}, which - using the Hong-Ou-Mandel effect - provides a optical implementation of a shallow neural network, using constant $\mathcal{O}(1)$ computational and optical resources with respect to its classical counterpart, an imaging apparatus combined with a classical neural network. We discuss how to obtain a multiclass model without modifying the Hong-Ou-Mandel architecture. To do so, we combine multiple quantum binary classifiers using strategies like one-vs-one (OvO), one-vs-rest (OvR), and a binary decision tree (DT), focusing on the computational overhead introduced by each protocol. In contrast to classical scenarios where the starting cost is already $\mathcal{O}(K)$ (so that an overhead of the same order is not relevant), in the quantum case the overhead is of particular importance to assess whether it kills the exponential speedup for a multiclass problem. At inference, we show that the decision tree introduces an overhead that is at most logarithmic in the number of classes, preserving the exponential speedup with respect to the input size (and to the number of hidden neurons for a shallow network). We compare these strategies against a classical model with higher expressive power, using three standard image classification benchmarks.

\section{METHOD}
In this section, we describe how to generalize a binary quantum classifier to the multiclass scenario, using classical post-processing methods like OvO, OvR, and DT. Consider the problem of classifying a dataset $D$ that contains $K$ classes. All three methods decompose the $K$-class problem into a collection of binary tasks, each solved by an independent quantum classifier, i.e.~separately trained and evaluated. The methods differ in the decomposition type, and in how the binary outputs are combined into a multinomial prediction. Being a hybrid (quantum-classical) implementation, we highlight the computational overhead introduced by each method at inference (namely, after training).
\begin{figure}[t]
	\centering
	\includegraphics[width = 0.35\textwidth]{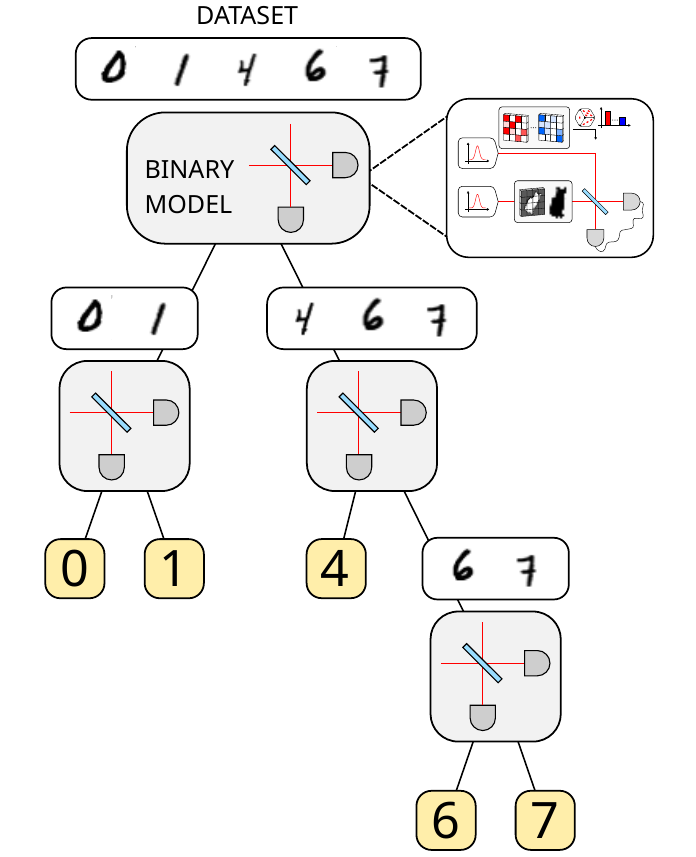}%
	\caption{\label{fig:Setup}Multinomial classifier implemented using a decision tree. Each step splits the classes into binary partitions until each leaf contains a single class. Each node is assigned to a binary classifier. The inset shows an example where the binary classifier is implemented optically, using the coincidence rate in the Hong-Ou-Mandel effect as a classification score (the same used in our simulations).}
\end{figure}

In OvO, each binary classifier is associated to a pair of distinct classes $(k,k')$ in $D$, with $k,k'\in\{0,\ldots,K-1 \, | \, k < k'\}$ and with model parameters $\theta_{kk'}$. In particular, objects of class $k$ and $k'$ are assigned to the label $0$ and $1$, respectively. After training, a test input $x$ is submitted to all classifiers in parallel, each casting a vote for one of the two possible outcomes. By thresholding the outputs at $0.5$, each vote is given to the class $k$ if $f_{\theta_{kk'}}(x) < 0.5$, or to the class $k'$ otherwise. The multinomial prediction $y$ is assigned by majority voting, with ties broken in favor of the smallest label. At inference, this method requires $K(K-1)/2$ binary classifiers. Asymptotically, it introduces a computational overhead of $\mathcal{O}(K^2)$ with respect to the binary case.

In OvR, one binary classifier is assigned to each class $k \in \{0,1,\ldots,K-1\}$, with parameters $\theta_k$: the $k$-th classifier is trained to distinguish $k$ from all the remaining classes in $\{0,\ldots,k-1,k+1,\ldots,K-1\}$. In particular, we assign label $1$ to the inputs belonging to the class $k$, and label $0$ to all others. Once all binary classifiers are trained, a test input $x$ is fed into each of them in parallel, yielding a score $f_{\theta_k}(x) \in [0,1]$ that represents the probability that $x$ belongs to the $k$-th class. The multinomial prediction is obtained by choosing the model with maximal score, namely $y = \argmax_{k} f_{\theta_k}(x)$. At inference, this method requires the evaluation of $K$ binary classifiers, i.e.~a computational overhead of $\mathcal{O}(K)$ with respect to the binary case.

Finally, the DT recursively splits the set of $K$ classes into binary partitions, organizing the dataset in subsets with a hierarchical structure. Its generation goes as follows. At each node, the set of input classes, i.e.~the output of the previous step, is split among the left and right branches. Recursion continues until each leaf (external node) contains a single class. Then, the $k$-th node is assigned to a binary classifier with parameters $\theta_k$. Each model controls the flow through the tree, and it is trained to distinguish the group of classes in the left and right branches, with labels $0$ and $1$, respectively. After training, a test input $x$ is fed into the root node and propagated through the tree. At each node, the input is directed to the left or right branch, depending on whether the classifier yields $f_{\theta_{k}}(x) < 0.5$ or $\geq 0.5$, respectively. The multinomial prediction is obtained as soon as a leaf is reached. See \cref{fig:Setup} for a schematic representation. Since this model relies on feedforward propagation, it can be only evaluated sequentially and - at the sam-e time - does not require the evaluation of all the models. At each level, only one node is interrogated, for a total computational overhead of $\mathcal{O}(\log_2K)$ at inference (exponentially lower than the number of nodes). The drawback is that, in principle, the classification will depend on which among all the possible trees is chosen, namely on which classes are in the left/right branches at each node. Surprisingly, as discussed below, our simulations suggest that there is no such dependence. As is customary, we do not consider the resource cost of the training stage, but just of the classification.

\section{RESULTS}
\begin{table*}[t]
\centering
\begin{minipage}{0.48\textwidth}
\centering
\begin{tabular}{lcccccc}
\toprule
 & \multicolumn{3}{c}{Quantum $(\%)$} & \multicolumn{3}{c}{Classical $(\%)$} \\
\cmidrule(lr){2-4} \cmidrule(lr){5-7}
Dataset & OvO & OvR & DT & OvO & OvR & DT \\
\midrule
MNIST   & $96.6$ & $97.5$ & $95.3$ & $98.7$ & $98.7$ & $98.3$ \\
Fashion & $90.6$ & $91.7$ & $89.6$ & $92.6$ & $93.2$ & $92.2$ \\
CIFAR-10 & $44.1$ & $45.1$ & $40.5$ & $47.8$ & $47.8$ & $44.1$ \\
\bottomrule
\end{tabular}
\caption{Average accuracy for a $6$-class problem using the first $6$ labels of the dataset, i.e.~$(0,1,\ldots,5)$, tested on the validation data using the one-vs-one (OvO), one-vs-rest (OvR) and binary decision tree (DT) methods. Both the quantum and the classical models are trained with $20$ neurons and $128$ samples per batch, and they show comparable accuracy (even though the quantum uses less resources).\label{tab:Table1}}
\end{minipage}
\hfill
\begin{minipage}{0.48\textwidth}
\centering
\begin{tabular}{lcccccc}
\toprule
 & \multicolumn{3}{c}{Quantum $(\%)$} & \multicolumn{3}{c}{Classical $(\%)$} \\
\cmidrule(lr){2-4} \cmidrule(lr){5-7}
Dataset & OvO & OvR & DT & OvO & OvR & DT \\
\midrule
MNIST   & $98.0$ & $98.9$ & $97.4$ & $99.2$ & $99.3$ & $99.1$ \\
Fashion & $93.1$ & $94.6$ & $92.3$ & $95.5$ & $95.8$ & $95.0$ \\
CIFAR-10 & $56.5$ & $57.3$ & $55.4$ & $62.6$ & $63.8$ & $62.2$ \\
\bottomrule
\end{tabular}
\caption{Average accuracy for a $4$-class problem using the first $4$ labels of the dataset, i.e.~$(0,1,2,3)$, tested on the validation data using the one-vs-one (OvO), one-vs-rest (OvR) and binary decision tree (DT) methods. Both the quantum and the classical models are trained with $40$ neurons and $64$ samples per batch, and again show comparable accuracy.\label{tab:Table2}}
\end{minipage}
\end{table*}

In this section, we present numerical simulations that benchmark the performance of the multinomial classifier, by piping multiple quantum binary models into the post-processing strategies reviewed in the previous section (OvO, OvR and DT). We compare them by substituting the quantum architecture with a classical baseline, obtained using the same procedure. For the quantum model, we consider a Hong-Ou-Mandel two-photon setup \cite{roncallo2025shallow} that uses the coincidence rate as score for binary classification. Here, the input data and the model parameters are encoded in the spectral amplitudes of the photon spatial modes, using a pure state and a density operator (i.e.~a mixed state of $M$ pure components), respectively. Detection is performed by two bucket detectors placed after the beamsplitter, which ignore any spatial resolution. As shown in~\citep{roncallo2025shallow}, this optical setup is mathematically equivalent to a shallow neural network of $M$ hidden neurons with square modulus activation function, followed by a linear output layer with bias and sigmoid activation. Due to the way parameters are encoded, the weights of each hidden neuron are constrained to be $L^2$-normalized (their square absolute values sum to one), while the weights of the output layer are constrained to be non-negative and $L^1$-normalized (they sum to one). Compared to its classical counterpart, this model provides a superexponential speedup, i.e.~at inference, it can classify a sample of $N$ features using constant $\mathcal{O}(1)$ computational and optical resources.

We simulate the quantum model directly in PyTorch, by leveraging its mathematical equivalence to a classical shallow network with $M$ hidden neurons and $N$ input features, then enforcing the $L^2$ and $L^1$ normalization constraints on the hidden and output layers, respectively. Such constraints can be relaxed, by multiplying the output by the $L^1$ normalization constant before the sigmoid activation function (this is not possible for the $L^2$ constraints, since keeping track of each hidden neuron normalization would introduce a computational overhead much larger than the advantage). The classical baseline shares the same structure but stacks multiple hidden layers with depth $D$ with $M$ neurons per layer (without constraints), using ReLU activation, batch normalization, and dropout after each layer. This architecture is capable of higher expressive power than its quantum counterpart. This choice is deliberate: testing against a more expressive model ensures that poor performance - when affecting both the quantum and classical models - can be attributed to a bottleneck in the post-processing strategies rather than to the architectural limitations of the quantum model. 

All the methods reviewed in the previous section combine multiple (quantum or classical) models to obtain a multinomial classifier. This means that each binary model needs to be separately trained to solve a binary task. The multinomial prediction is obtained at inference, without further retraining. Training is done using the binary cross-entropy as loss function, optimized with mini-batch gradient descent, learning rate $0.05$, momentum $0.09$ and decay $10^{-4}$. For our benchmarks, we consider three widely recognized datasets. The MNIST dataset contains $28 \times 28$ greyscale images of handwritten digits, distributed across $10$ classes. The Fashion MNIST dataset contains $28 \times 28$ greyscale images of clothing items, across $10$ classes. The CIFAR-10 dataset consists of $32 \times 32$ color images, distributed across $10$ classes. These datasets progressively represent a harder benchmark, due to the complexity of their visual patterns.  We simplify the task by considering a subset of the target labels, i.e.~$K<10$. Each dataset is pre-processed before being fed into the models; it undergoes greyscale conversion (for the CIFAR-10), flattening, standardization, then sample-wise normalization with respect to the $L^2$ norm (so that it can be interpreted as the discretized spectrum of a single-photon state). 

Results for the $6$-class and $4$-class problems are reported in \cref{tab:Table1,tab:Table2}, respectively. As figure of merit, we consider the accuracy defined as the ratio of correctly classified inputs over the total number of samples per class, uniformly averaged over all $K$ classes. Across all datasets and strategies, the quantum and classical models achieve comparable accuracy, with the classical models showing a small but consistent advantage. We interpret this as a consequence of the normalization constraints, which reduce the expressivity of the quantum models. Importantly, the choice of multiclass strategy has negligible influence on the final accuracy: OvO, OvR, and DT show similar performances in all scenarios. This is a key observation. These strategies introduce significantly different overhead, $\mathcal{O}(K^2)$, $\mathcal{O}(K)$, and $\mathcal{O}(\log_2K)$, respectively. Since this does not lead to any accuracy difference, the decision tree represents the most efficient choice: it is possible to obtain a multinomial classifier from a collection of binary models, by introducing at most a logarithmic overhead. 

An entry-wise comparison shows that \cref{tab:Table1} has lower accuracies than \cref{tab:Table2}, overall. We investigate this behaviour by measuring the accuracy of the DT method for different $K$ (from $2$ to $10$), both for the quantum and the classical models while keeping the architecture fixed. Results are reported in \cref{tab:Table3}. As $K$ increases (up to $10$), the averaged accuracy decreases monotonically for both the quantum and classical models. This degradation is consistent with the complexity scaling of the problem, and maintains an accuracy gap of $30-23\%$ and $35-28\%$ with respect to the random guess, respectively for the quantum and the classical models.

Among the three strategies, DT is the only one that requires feedforward propagation (while OvO and OvR can be executed in parallel). Misclassified samples at the root node propagate through the entire tree. Such a node contains also the largest and most heterogeneous partition of classes; it is the one most likely to produce errors. We consider additional tests on DT, to assess whether this is the bottleneck and to what extent it can be mitigated. In the first test, we consider a fixed tree topology, while we train the binary model at the root node with $3$ different random seeds. The prediction (i.e.~which node needs to be evaluated after) is obtained by majority voting. In the second test, $3$ different topologies are generated using different random partitions, each independently trained. At inference, each tree produces a full multinomial prediction; the final label is assigned by majority voting across the predicted leaves (with ties broken in favor of the first run). These tests introduce an overhead of $\mathcal{O}(1)$ and $\mathcal{O}(3)$, respectively at the root and tree level. Our simulations show that neither strategy leads to a consistent accuracy improvement. In the former case, this suggests that variance in training does not produce any noticeable difference (provided that the optimization schedule has converged). In the latter case, it suggests that tree topology and the partition choice are not a bottleneck of the model.

\begin{table}[th]
\centering
\begin{tabular}{lcccccccccc}
\toprule
$K$       & 2 & 3 & 4 & 5 & 6 & 7 & 8 & 9 & 10 \\
\midrule
Quantum $(\%)$   & $78.4$ & $66.2$ & $57.0$ & $47.6$ & $40.2$ & $38.3$ & $36.5$ & $34.4$ & $33.2$ \\
Classical $(\%)$ & $85.0$ & $70.9$ & $60.2$ & $50.0$ & $44.8$ & $40.6$ & $41.1$ & $40.3$ & $38.0$ \\
Random $(\%)$    & $50.0$ & $33.3$ & $25.0$ & $20.0$ & $16.7$ & $14.3$ & $12.5$ & $11.1$ & $10.0$ \\
\bottomrule
\end{tabular}
\caption{Average accuracy for a $K$-class problem with labels $(0,1,\ldots,K-1)$, tested on the CIFAR-10 validation data using the binary decision tree method, both for the quantum and the classical models, showing the decrease of the accuracy for growing $K$.\label{tab:Table3}
}
\end{table}

\section{CONCLUSIONS}
We considered the problem of using quantum binary models to solve multiclass problems. Without modifying the underlying architecture, we combined multiple models in post-processing, using techniques like one-vs-one, one-vs-rest and a binary decision tree. We benchmarked their performance, specializing to our Hong-Ou-Mandel setup and comparing against a classical binary model (generalized using the same approach). 

The choice of the strategy has a negligible impact on the average accuracy. In contrast, different strategies lead to different computational overheads. The decision tree represents the cheapest implementation, being based on a sequential (rather than parallel) implementation, which bypasses the evaluation of all the nodes. Asymptotically, it requires $\mathcal{O}(\log_2K)$ evaluations for a $K$-class problem. Combined with the constant scaling $\mathcal{O}(1)$ of the Hong-Ou-Mandel classifier, this shows that it is possible to achieve a quantum multinomial classifier with exponential speedup.

As shown by our simulations, increasing the complexity of the task, i.e.~the number of classes to predict, leads to a progressive degradation of the average accuracy. This affects both the quantum and the classical models, suggesting that the bottleneck is represented by the post-processing strategy rather than by the architectural choice. Eventually, improving the model requires adding the decision tree to the training pipeline, rather than treating it as a post-processing step. Such joint training procedure could coherently exploit the correlations across classes, potentially improving accuracy against agnostic partitioning alone.

\section*{ACKNOWLEDGEMENTS}
S.R. acknowledges support from the PRIN MUR Project 2022SW3RPY. A.R.M. acknowledges support from the PNRR MUR Project PE0000023-NQSTI. C.M. acknowledges support from the National Research Centre for HPC, Big Data and Quantum Computing, PNRR MUR Project CN0000013-ICSC. L.M. acknowledges support from the PRIN MUR Project 2022RATBS4. SL: This material is based upon work supported by, or in part by, the U. S. Army Research Laboratory and the U. S. Army Research Office under contract/grant number W911NF2310255, and by DoE under contract, DE-SC0012704.

\section*{CODE AVAILABILITY}
The underlying code that generated the data for this study is openly available on GitHub~\citep{rep:QON}.





\FloatBarrier
\bibliography{refs.bib}

\begin{thebibliography}{32}%
\makeatletter
\providecommand \@ifxundefined [1]{%
 \@ifx{#1\undefined}
}%
\providecommand \@ifnum [1]{%
 \ifnum #1\expandafter \@firstoftwo
 \else \expandafter \@secondoftwo
 \fi
}%
\providecommand \@ifx [1]{%
 \ifx #1\expandafter \@firstoftwo
 \else \expandafter \@secondoftwo
 \fi
}%
\providecommand \natexlab [1]{#1}%
\providecommand \enquote  [1]{``#1''}%
\providecommand \bibnamefont  [1]{#1}%
\providecommand \bibfnamefont [1]{#1}%
\providecommand \citenamefont [1]{#1}%
\providecommand \href@noop [0]{\@secondoftwo}%
\providecommand \href [0]{\begingroup \@sanitize@url \@href}%
\providecommand \@href[1]{\@@startlink{#1}\@@href}%
\providecommand \@@href[1]{\endgroup#1\@@endlink}%
\providecommand \@sanitize@url [0]{\catcode `\\12\catcode `\$12\catcode
  `\&12\catcode `\#12\catcode `\^12\catcode `\_12\catcode `\%12\relax}%
\providecommand \@@startlink[1]{}%
\providecommand \@@endlink[0]{}%
\providecommand \url  [0]{\begingroup\@sanitize@url \@url }%
\providecommand \@url [1]{\endgroup\@href {#1}{\urlprefix }}%
\providecommand \urlprefix  [0]{URL }%
\providecommand \Eprint [0]{\href }%
\providecommand \doibase [0]{https://doi.org/}%
\providecommand \selectlanguage [0]{\@gobble}%
\providecommand \bibinfo  [0]{\@secondoftwo}%
\providecommand \bibfield  [0]{\@secondoftwo}%
\providecommand \translation [1]{[#1]}%
\providecommand \BibitemOpen [0]{}%
\providecommand \bibitemStop [0]{}%
\providecommand \bibitemNoStop [0]{.\EOS\space}%
\providecommand \EOS [0]{\spacefactor3000\relax}%
\providecommand \BibitemShut  [1]{\csname bibitem#1\endcsname}%
\let\auto@bib@innerbib\@empty
\bibitem [{\citenamefont {Goodfellow}\ \emph {et~al.}(2016)\citenamefont
  {Goodfellow}, \citenamefont {Bengio},\ and\ \citenamefont
  {Courville}}]{book:Goodfellow}%
  \BibitemOpen
  \bibfield  {author} {\bibinfo {author} {\bibfnamefont {I.}~\bibnamefont
  {Goodfellow}}, \bibinfo {author} {\bibfnamefont {Y.}~\bibnamefont {Bengio}},\
  and\ \bibinfo {author} {\bibfnamefont {A.}~\bibnamefont {Courville}},\
  }\href@noop {} {\emph {\bibinfo {title} {Deep Learning}}}\ (\bibinfo
  {publisher} {MIT Press},\ \bibinfo {year} {2016})\ \bibinfo {note}
  {\url{http://www.deeplearningbook.org}}\BibitemShut {NoStop}%
\bibitem [{\citenamefont {Lloyd}\ \emph {et~al.}(2013)\citenamefont {Lloyd},
  \citenamefont {Mohseni},\ and\ \citenamefont
  {Rebentrost}}]{lloyd2013quantum}%
  \BibitemOpen
  \bibfield  {author} {\bibinfo {author} {\bibfnamefont {S.}~\bibnamefont
  {Lloyd}}, \bibinfo {author} {\bibfnamefont {M.}~\bibnamefont {Mohseni}},\
  and\ \bibinfo {author} {\bibfnamefont {P.}~\bibnamefont {Rebentrost}},\
  }\href@noop {} {\bibinfo {title} {Quantum algorithms for supervised and
  unsupervised machine learning}} (\bibinfo {year} {2013}),\ \Eprint
  {https://arxiv.org/abs/1307.0411} {arXiv:1307.0411 [quant-ph]} \BibitemShut
  {NoStop}%
\bibitem [{\citenamefont {Cai}\ \emph {et~al.}(2015)\citenamefont {Cai},
  \citenamefont {Wu}, \citenamefont {Su}, \citenamefont {Chen}, \citenamefont
  {Wang}, \citenamefont {Li}, \citenamefont {Liu}, \citenamefont {Lu},\ and\
  \citenamefont {Pan}}]{cai2015entanglement}%
  \BibitemOpen
  \bibfield  {author} {\bibinfo {author} {\bibfnamefont {X.-D.}\ \bibnamefont
  {Cai}}, \bibinfo {author} {\bibfnamefont {D.}~\bibnamefont {Wu}}, \bibinfo
  {author} {\bibfnamefont {Z.-E.}\ \bibnamefont {Su}}, \bibinfo {author}
  {\bibfnamefont {M.-C.}\ \bibnamefont {Chen}}, \bibinfo {author}
  {\bibfnamefont {X.-L.}\ \bibnamefont {Wang}}, \bibinfo {author}
  {\bibfnamefont {L.}~\bibnamefont {Li}}, \bibinfo {author} {\bibfnamefont
  {N.-L.}\ \bibnamefont {Liu}}, \bibinfo {author} {\bibfnamefont {C.-Y.}\
  \bibnamefont {Lu}},\ and\ \bibinfo {author} {\bibfnamefont {J.-W.}\
  \bibnamefont {Pan}},\ }\bibfield  {title} {\bibinfo {title}
  {Entanglement-based machine learning on a quantum computer},\ }\href
  {https://doi.org/https://doi.org/10.1103/PhysRevLett.114.110504} {\bibfield
  {journal} {\bibinfo  {journal} {Phys. Rev. Lett.}\ }\textbf {\bibinfo
  {volume} {114}},\ \bibinfo {pages} {110504} (\bibinfo {year}
  {2015})}\BibitemShut {NoStop}%
\bibitem [{\citenamefont {Tacchino}\ \emph {et~al.}(2019)\citenamefont
  {Tacchino}, \citenamefont {Macchiavello}, \citenamefont {Gerace},\ and\
  \citenamefont {Bajoni}}]{tacchino2019artificial}%
  \BibitemOpen
  \bibfield  {author} {\bibinfo {author} {\bibfnamefont {F.}~\bibnamefont
  {Tacchino}}, \bibinfo {author} {\bibfnamefont {C.}~\bibnamefont
  {Macchiavello}}, \bibinfo {author} {\bibfnamefont {D.}~\bibnamefont
  {Gerace}},\ and\ \bibinfo {author} {\bibfnamefont {D.}~\bibnamefont
  {Bajoni}},\ }\bibfield  {title} {\bibinfo {title} {An artificial neuron
  implemented on an actual quantum processor},\ }\href
  {https://doi.org/https://doi.org/10.1038/s41534-019-0140-4} {\bibfield
  {journal} {\bibinfo  {journal} {Npj Quantum Inf.}\ }\textbf {\bibinfo
  {volume} {5}},\ \bibinfo {pages} {26} (\bibinfo {year} {2019})}\BibitemShut
  {NoStop}%
\bibitem [{\citenamefont {Benatti}\ \emph {et~al.}(2019)\citenamefont
  {Benatti}, \citenamefont {Mancini},\ and\ \citenamefont
  {Mangini}}]{benatti2019continuous}%
  \BibitemOpen
  \bibfield  {author} {\bibinfo {author} {\bibfnamefont {F.}~\bibnamefont
  {Benatti}}, \bibinfo {author} {\bibfnamefont {S.}~\bibnamefont {Mancini}},\
  and\ \bibinfo {author} {\bibfnamefont {S.}~\bibnamefont {Mangini}},\
  }\bibfield  {title} {\bibinfo {title} {Continuous variable quantum
  perceptron},\ }\href
  {https://doi.org/https://doi.org/10.1142/s0219749919410090} {\bibfield
  {journal} {\bibinfo  {journal} {Int. J. Quantum Inf.}\ }\textbf {\bibinfo
  {volume} {17}},\ \bibinfo {pages} {1941009} (\bibinfo {year}
  {2019})}\BibitemShut {NoStop}%
\bibitem [{\citenamefont {Mangini}\ \emph {et~al.}(2020)\citenamefont
  {Mangini}, \citenamefont {Tacchino}, \citenamefont {Gerace}, \citenamefont
  {Macchiavello},\ and\ \citenamefont {Bajoni}}]{mangini2020quantum}%
  \BibitemOpen
  \bibfield  {author} {\bibinfo {author} {\bibfnamefont {S.}~\bibnamefont
  {Mangini}}, \bibinfo {author} {\bibfnamefont {F.}~\bibnamefont {Tacchino}},
  \bibinfo {author} {\bibfnamefont {D.}~\bibnamefont {Gerace}}, \bibinfo
  {author} {\bibfnamefont {C.}~\bibnamefont {Macchiavello}},\ and\ \bibinfo
  {author} {\bibfnamefont {D.}~\bibnamefont {Bajoni}},\ }\bibfield  {title}
  {\bibinfo {title} {Quantum computing model of an artificial neuron with
  continuously valued input data},\ }\href
  {https://doi.org/https://doi.org/10.1088/2632-2153/abaf98} {\bibfield
  {journal} {\bibinfo  {journal} {Mach. Learn.: Sci. Technol.}\ }\textbf
  {\bibinfo {volume} {1}},\ \bibinfo {pages} {045008} (\bibinfo {year}
  {2020})}\BibitemShut {NoStop}%
\bibitem [{\citenamefont {Steinbrecher}\ \emph {et~al.}(2019)\citenamefont
  {Steinbrecher}, \citenamefont {Olson}, \citenamefont {Englund},\ and\
  \citenamefont {Carolan}}]{steinbrecher2019quantum}%
  \BibitemOpen
  \bibfield  {author} {\bibinfo {author} {\bibfnamefont {G.~R.}\ \bibnamefont
  {Steinbrecher}}, \bibinfo {author} {\bibfnamefont {J.~P.}\ \bibnamefont
  {Olson}}, \bibinfo {author} {\bibfnamefont {D.}~\bibnamefont {Englund}},\
  and\ \bibinfo {author} {\bibfnamefont {J.}~\bibnamefont {Carolan}},\
  }\bibfield  {title} {\bibinfo {title} {Quantum optical neural networks},\
  }\href {https://doi.org/https://doi.org/10.1038/s41534-019-0174-7} {\bibfield
   {journal} {\bibinfo  {journal} {Npj Quantum Inf.}\ }\textbf {\bibinfo
  {volume} {5}},\ \bibinfo {pages} {60} (\bibinfo {year} {2019})}\BibitemShut
  {NoStop}%
\bibitem [{\citenamefont {Killoran}\ \emph {et~al.}(2019)\citenamefont
  {Killoran}, \citenamefont {Bromley}, \citenamefont {Arrazola}, \citenamefont
  {Schuld}, \citenamefont {Quesada},\ and\ \citenamefont
  {Lloyd}}]{killoran2019continuous}%
  \BibitemOpen
  \bibfield  {author} {\bibinfo {author} {\bibfnamefont {N.}~\bibnamefont
  {Killoran}}, \bibinfo {author} {\bibfnamefont {T.~R.}\ \bibnamefont
  {Bromley}}, \bibinfo {author} {\bibfnamefont {J.~M.}\ \bibnamefont
  {Arrazola}}, \bibinfo {author} {\bibfnamefont {M.}~\bibnamefont {Schuld}},
  \bibinfo {author} {\bibfnamefont {N.}~\bibnamefont {Quesada}},\ and\ \bibinfo
  {author} {\bibfnamefont {S.}~\bibnamefont {Lloyd}},\ }\bibfield  {title}
  {\bibinfo {title} {Continuous-variable quantum neural networks},\ }\href
  {https://doi.org/https://doi.org/10.1103/PhysRevResearch.1.033063} {\bibfield
   {journal} {\bibinfo  {journal} {Phys. Rev. Res.}\ }\textbf {\bibinfo
  {volume} {1}},\ \bibinfo {pages} {033063} (\bibinfo {year}
  {2019})}\BibitemShut {NoStop}%
\bibitem [{\citenamefont {Sui}\ \emph {et~al.}(2020)\citenamefont {Sui},
  \citenamefont {Wu}, \citenamefont {Liu}, \citenamefont {Chen},\ and\
  \citenamefont {Gu}}]{sui2020review}%
  \BibitemOpen
  \bibfield  {author} {\bibinfo {author} {\bibfnamefont {X.}~\bibnamefont
  {Sui}}, \bibinfo {author} {\bibfnamefont {Q.}~\bibnamefont {Wu}}, \bibinfo
  {author} {\bibfnamefont {J.}~\bibnamefont {Liu}}, \bibinfo {author}
  {\bibfnamefont {Q.}~\bibnamefont {Chen}},\ and\ \bibinfo {author}
  {\bibfnamefont {G.}~\bibnamefont {Gu}},\ }\bibfield  {title} {\bibinfo
  {title} {A review of optical neural networks},\ }\href
  {https://doi.org/https://doi.org/10.1109/ACCESS.2020.2987333} {\bibfield
  {journal} {\bibinfo  {journal} {IEEE Access}\ }\textbf {\bibinfo {volume}
  {8}},\ \bibinfo {pages} {70773} (\bibinfo {year} {2020})}\BibitemShut
  {NoStop}%
\bibitem [{\citenamefont {Zhang}\ \emph {et~al.}(2021)\citenamefont {Zhang},
  \citenamefont {Zhan}, \citenamefont {Liao}, \citenamefont {Zheng},
  \citenamefont {Jiang}, \citenamefont {Mi}, \citenamefont {Yao},\ and\
  \citenamefont {Zhang}}]{zhang2021quantum}%
  \BibitemOpen
  \bibfield  {author} {\bibinfo {author} {\bibfnamefont {A.}~\bibnamefont
  {Zhang}}, \bibinfo {author} {\bibfnamefont {H.}~\bibnamefont {Zhan}},
  \bibinfo {author} {\bibfnamefont {J.}~\bibnamefont {Liao}}, \bibinfo {author}
  {\bibfnamefont {K.}~\bibnamefont {Zheng}}, \bibinfo {author} {\bibfnamefont
  {T.}~\bibnamefont {Jiang}}, \bibinfo {author} {\bibfnamefont
  {M.}~\bibnamefont {Mi}}, \bibinfo {author} {\bibfnamefont {P.}~\bibnamefont
  {Yao}},\ and\ \bibinfo {author} {\bibfnamefont {L.}~\bibnamefont {Zhang}},\
  }\bibfield  {title} {\bibinfo {title} {Quantum verification of {NP} problems
  with single photons and linear optics},\ }\href
  {https://doi.org/https://doi.org/10.1038/s41377-021-00608-4} {\bibfield
  {journal} {\bibinfo  {journal} {Light Sci. Appl.}\ }\textbf {\bibinfo
  {volume} {10}},\ \bibinfo {pages} {169} (\bibinfo {year} {2021})}\BibitemShut
  {NoStop}%
\bibitem [{\citenamefont {Spall}\ \emph {et~al.}(2022)\citenamefont {Spall},
  \citenamefont {Guo},\ and\ \citenamefont {Lvovsky}}]{spall2022hybrid}%
  \BibitemOpen
  \bibfield  {author} {\bibinfo {author} {\bibfnamefont {J.}~\bibnamefont
  {Spall}}, \bibinfo {author} {\bibfnamefont {X.}~\bibnamefont {Guo}},\ and\
  \bibinfo {author} {\bibfnamefont {A.~I.}\ \bibnamefont {Lvovsky}},\
  }\bibfield  {title} {\bibinfo {title} {Hybrid training of optical neural
  networks},\ }\href {https://doi.org/https://doi.org/10.1364/OPTICA.456108}
  {\bibfield  {journal} {\bibinfo  {journal} {Optica}\ }\textbf {\bibinfo
  {volume} {9}},\ \bibinfo {pages} {803} (\bibinfo {year} {2022})}\BibitemShut
  {NoStop}%
\bibitem [{\citenamefont {Stanev}\ \emph {et~al.}(2023)\citenamefont {Stanev},
  \citenamefont {Spagnolo},\ and\ \citenamefont
  {Sciarrino}}]{stanev2023deterministic}%
  \BibitemOpen
  \bibfield  {author} {\bibinfo {author} {\bibfnamefont {D.}~\bibnamefont
  {Stanev}}, \bibinfo {author} {\bibfnamefont {N.}~\bibnamefont {Spagnolo}},\
  and\ \bibinfo {author} {\bibfnamefont {F.}~\bibnamefont {Sciarrino}},\
  }\bibfield  {title} {\bibinfo {title} {Deterministic optimal quantum cloning
  via a quantum-optical neural network},\ }\href
  {https://doi.org/https://doi.org/10.1103/PhysRevResearch.5.013139} {\bibfield
   {journal} {\bibinfo  {journal} {Phys. Rev. Res.}\ }\textbf {\bibinfo
  {volume} {5}},\ \bibinfo {pages} {013139} (\bibinfo {year}
  {2023})}\BibitemShut {NoStop}%
\bibitem [{\citenamefont {Wood}\ \emph {et~al.}(2024)\citenamefont {Wood},
  \citenamefont {Shrapnel},\ and\ \citenamefont {Milburn}}]{wood2024kerr}%
  \BibitemOpen
  \bibfield  {author} {\bibinfo {author} {\bibfnamefont {C.}~\bibnamefont
  {Wood}}, \bibinfo {author} {\bibfnamefont {S.}~\bibnamefont {Shrapnel}},\
  and\ \bibinfo {author} {\bibfnamefont {G.~J.}\ \bibnamefont {Milburn}},\
  }\href@noop {} {\bibinfo {title} {A {K}err kernel quantum learning machine}}
  (\bibinfo {year} {2024}),\ \Eprint {https://arxiv.org/abs/2404.01787}
  {arXiv:2404.01787 [quant-ph]} \BibitemShut {NoStop}%
\bibitem [{\citenamefont {Spall}\ \emph {et~al.}(2025)\citenamefont {Spall},
  \citenamefont {Guo},\ and\ \citenamefont {Lvovsky}}]{spall2025training}%
  \BibitemOpen
  \bibfield  {author} {\bibinfo {author} {\bibfnamefont {J.}~\bibnamefont
  {Spall}}, \bibinfo {author} {\bibfnamefont {X.}~\bibnamefont {Guo}},\ and\
  \bibinfo {author} {\bibfnamefont {A.~I.}\ \bibnamefont {Lvovsky}},\
  }\bibfield  {title} {\bibinfo {title} {Training neural networks with
  end-to-end optical backpropagation},\ }\href
  {https://doi.org/https://doi.org/10.1117/1.AP.7.1.016004} {\bibfield
  {journal} {\bibinfo  {journal} {Adv. Photonics}\ }\textbf {\bibinfo {volume}
  {7}},\ \bibinfo {pages} {016004} (\bibinfo {year} {2025})}\BibitemShut
  {NoStop}%
\bibitem [{\citenamefont {Hoch}\ \emph {et~al.}(2025)\citenamefont {Hoch},
  \citenamefont {Caruccio}, \citenamefont {Rodari}, \citenamefont
  {Francalanci}, \citenamefont {Suprano}, \citenamefont {Giordani},
  \citenamefont {Carvacho}, \citenamefont {Spagnolo}, \citenamefont {Koudia},
  \citenamefont {Proietti} \emph {et~al.}}]{hoch2025quantum}%
  \BibitemOpen
  \bibfield  {author} {\bibinfo {author} {\bibfnamefont {F.}~\bibnamefont
  {Hoch}}, \bibinfo {author} {\bibfnamefont {E.}~\bibnamefont {Caruccio}},
  \bibinfo {author} {\bibfnamefont {G.}~\bibnamefont {Rodari}}, \bibinfo
  {author} {\bibfnamefont {T.}~\bibnamefont {Francalanci}}, \bibinfo {author}
  {\bibfnamefont {A.}~\bibnamefont {Suprano}}, \bibinfo {author} {\bibfnamefont
  {T.}~\bibnamefont {Giordani}}, \bibinfo {author} {\bibfnamefont
  {G.}~\bibnamefont {Carvacho}}, \bibinfo {author} {\bibfnamefont
  {N.}~\bibnamefont {Spagnolo}}, \bibinfo {author} {\bibfnamefont
  {S.}~\bibnamefont {Koudia}}, \bibinfo {author} {\bibfnamefont
  {M.}~\bibnamefont {Proietti}}, \emph {et~al.},\ }\bibfield  {title} {\bibinfo
  {title} {Quantum machine learning with adaptive boson sampling via
  post-selection},\ }\href
  {https://doi.org/https://doi.org/10.1038/s41467-025-55877-z} {\bibfield
  {journal} {\bibinfo  {journal} {Nat. Commun.}\ }\textbf {\bibinfo {volume}
  {16}},\ \bibinfo {pages} {902} (\bibinfo {year} {2025})}\BibitemShut
  {NoStop}%
\bibitem [{\citenamefont {Slabbert}\ and\ \citenamefont
  {Petruccione}(2025)}]{slabbert2025classical}%
  \BibitemOpen
  \bibfield  {author} {\bibinfo {author} {\bibfnamefont {D.}~\bibnamefont
  {Slabbert}}\ and\ \bibinfo {author} {\bibfnamefont {F.}~\bibnamefont
  {Petruccione}},\ }\bibfield  {title} {\bibinfo {title} {Classical-quantum
  approach to image classification: Autoencoders and quantum {SVM}s},\ }\href
  {https://doi.org/https://doi.org/10.1116/5.0261885} {\bibfield  {journal}
  {\bibinfo  {journal} {AVS Quantum Sci.}\ }\textbf {\bibinfo {volume} {7}},\
  \bibinfo {pages} {023804} (\bibinfo {year} {2025})}\BibitemShut {NoStop}%
\bibitem [{\citenamefont {Sun}\ \emph {et~al.}(2025)\citenamefont {Sun},
  \citenamefont {Li}, \citenamefont {Xiang}, \citenamefont {Yuan},
  \citenamefont {Hu}, \citenamefont {Hua}, \citenamefont {Jiang}, \citenamefont
  {Zhu},\ and\ \citenamefont {Fu}}]{sun2025scalable}%
  \BibitemOpen
  \bibfield  {author} {\bibinfo {author} {\bibfnamefont {Y.}~\bibnamefont
  {Sun}}, \bibinfo {author} {\bibfnamefont {D.}~\bibnamefont {Li}}, \bibinfo
  {author} {\bibfnamefont {Q.}~\bibnamefont {Xiang}}, \bibinfo {author}
  {\bibfnamefont {Y.}~\bibnamefont {Yuan}}, \bibinfo {author} {\bibfnamefont
  {Z.}~\bibnamefont {Hu}}, \bibinfo {author} {\bibfnamefont {X.}~\bibnamefont
  {Hua}}, \bibinfo {author} {\bibfnamefont {Y.}~\bibnamefont {Jiang}}, \bibinfo
  {author} {\bibfnamefont {Y.}~\bibnamefont {Zhu}},\ and\ \bibinfo {author}
  {\bibfnamefont {Y.}~\bibnamefont {Fu}},\ }\bibfield  {title} {\bibinfo
  {title} {Scalable quantum convolutional neural network for image
  classification},\ }\href
  {https://doi.org/https://doi.org/10.1016/j.physa.2024.130226} {\bibfield
  {journal} {\bibinfo  {journal} {Phys. A: Stat. Mech. Appl.}\ }\textbf
  {\bibinfo {volume} {657}},\ \bibinfo {pages} {130226} (\bibinfo {year}
  {2025})}\BibitemShut {NoStop}%
\bibitem [{\citenamefont {Sakurai}\ \emph {et~al.}(2025)\citenamefont
  {Sakurai}, \citenamefont {Hayashi}, \citenamefont {Munro},\ and\
  \citenamefont {Nemoto}}]{sakurai2025quantum}%
  \BibitemOpen
  \bibfield  {author} {\bibinfo {author} {\bibfnamefont {A.}~\bibnamefont
  {Sakurai}}, \bibinfo {author} {\bibfnamefont {A.}~\bibnamefont {Hayashi}},
  \bibinfo {author} {\bibfnamefont {W.~J.}\ \bibnamefont {Munro}},\ and\
  \bibinfo {author} {\bibfnamefont {K.}~\bibnamefont {Nemoto}},\ }\bibfield
  {title} {\bibinfo {title} {Quantum optical reservoir computing powered by
  boson sampling},\ }\href
  {https://doi.org/https://doi.org/10.1364/OPTICAQ.541432} {\bibfield
  {journal} {\bibinfo  {journal} {Opt. Quantum}\ }\textbf {\bibinfo {volume}
  {3}},\ \bibinfo {pages} {238} (\bibinfo {year} {2025})}\BibitemShut {NoStop}%
\bibitem [{\citenamefont {Schuld}\ \emph {et~al.}(2020)\citenamefont {Schuld},
  \citenamefont {Bocharov}, \citenamefont {Svore},\ and\ \citenamefont
  {Wiebe}}]{schuld2020circuit}%
  \BibitemOpen
  \bibfield  {author} {\bibinfo {author} {\bibfnamefont {M.}~\bibnamefont
  {Schuld}}, \bibinfo {author} {\bibfnamefont {A.}~\bibnamefont {Bocharov}},
  \bibinfo {author} {\bibfnamefont {K.~M.}\ \bibnamefont {Svore}},\ and\
  \bibinfo {author} {\bibfnamefont {N.}~\bibnamefont {Wiebe}},\ }\bibfield
  {title} {\bibinfo {title} {Circuit-centric quantum classifiers},\ }\href
  {https://doi.org/https://doi.org/10.1103/PhysRevA.101.032308} {\bibfield
  {journal} {\bibinfo  {journal} {Phys. Rev. A}\ }\textbf {\bibinfo {volume}
  {101}},\ \bibinfo {pages} {032308} (\bibinfo {year} {2020})}\BibitemShut
  {NoStop}%
\bibitem [{\citenamefont {Havl{\'\i}{\v{c}}ek}\ \emph
  {et~al.}(2019)\citenamefont {Havl{\'\i}{\v{c}}ek}, \citenamefont
  {C{\'o}rcoles}, \citenamefont {Temme}, \citenamefont {Harrow}, \citenamefont
  {Kandala}, \citenamefont {Chow},\ and\ \citenamefont
  {Gambetta}}]{havlivcek2019supervised}%
  \BibitemOpen
  \bibfield  {author} {\bibinfo {author} {\bibfnamefont {V.}~\bibnamefont
  {Havl{\'\i}{\v{c}}ek}}, \bibinfo {author} {\bibfnamefont {A.~D.}\
  \bibnamefont {C{\'o}rcoles}}, \bibinfo {author} {\bibfnamefont
  {K.}~\bibnamefont {Temme}}, \bibinfo {author} {\bibfnamefont {A.~W.}\
  \bibnamefont {Harrow}}, \bibinfo {author} {\bibfnamefont {A.}~\bibnamefont
  {Kandala}}, \bibinfo {author} {\bibfnamefont {J.~M.}\ \bibnamefont {Chow}},\
  and\ \bibinfo {author} {\bibfnamefont {J.~M.}\ \bibnamefont {Gambetta}},\
  }\bibfield  {title} {\bibinfo {title} {Supervised learning with
  quantum-enhanced feature spaces},\ }\href
  {https://doi.org/https://doi.org/10.1038/s41586-019-0980-2} {\bibfield
  {journal} {\bibinfo  {journal} {Nature}\ }\textbf {\bibinfo {volume} {567}},\
  \bibinfo {pages} {209} (\bibinfo {year} {2019})}\BibitemShut {NoStop}%
\bibitem [{\citenamefont {Maheshwari}\ \emph {et~al.}(2021)\citenamefont
  {Maheshwari}, \citenamefont {Sierra-Sosa},\ and\ \citenamefont
  {Garcia-Zapirain}}]{maheshwari2021variational}%
  \BibitemOpen
  \bibfield  {author} {\bibinfo {author} {\bibfnamefont {D.}~\bibnamefont
  {Maheshwari}}, \bibinfo {author} {\bibfnamefont {D.}~\bibnamefont
  {Sierra-Sosa}},\ and\ \bibinfo {author} {\bibfnamefont {B.}~\bibnamefont
  {Garcia-Zapirain}},\ }\bibfield  {title} {\bibinfo {title} {Variational
  quantum classifier for binary classification: Real vs synthetic dataset},\
  }\href {https://doi.org/https://doi.org/10.1109/ACCESS.2021.3139323}
  {\bibfield  {journal} {\bibinfo  {journal} {IEEE access}\ }\textbf {\bibinfo
  {volume} {10}},\ \bibinfo {pages} {3705} (\bibinfo {year}
  {2021})}\BibitemShut {NoStop}%
\bibitem [{\citenamefont {Rebentrost}\ \emph {et~al.}(2014)\citenamefont
  {Rebentrost}, \citenamefont {Mohseni},\ and\ \citenamefont
  {Lloyd}}]{rebentrost2014quantum}%
  \BibitemOpen
  \bibfield  {author} {\bibinfo {author} {\bibfnamefont {P.}~\bibnamefont
  {Rebentrost}}, \bibinfo {author} {\bibfnamefont {M.}~\bibnamefont
  {Mohseni}},\ and\ \bibinfo {author} {\bibfnamefont {S.}~\bibnamefont
  {Lloyd}},\ }\bibfield  {title} {\bibinfo {title} {Quantum support vector
  machine for big data classification},\ }\href
  {https://doi.org/https://doi.org/10.1103/PhysRevLett.113.130503} {\bibfield
  {journal} {\bibinfo  {journal} {Phys. Rev. Lett.}\ }\textbf {\bibinfo
  {volume} {113}},\ \bibinfo {pages} {130503} (\bibinfo {year}
  {2014})}\BibitemShut {NoStop}%
\bibitem [{\citenamefont {Schuld}\ and\ \citenamefont
  {Killoran}(2019)}]{schuld2019quantum}%
  \BibitemOpen
  \bibfield  {author} {\bibinfo {author} {\bibfnamefont {M.}~\bibnamefont
  {Schuld}}\ and\ \bibinfo {author} {\bibfnamefont {N.}~\bibnamefont
  {Killoran}},\ }\bibfield  {title} {\bibinfo {title} {Quantum machine learning
  in feature hilbert spaces},\ }\href
  {https://doi.org/https://doi.org/10.1103/PhysRevLett.122.040504} {\bibfield
  {journal} {\bibinfo  {journal} {Phys. Rev. Lett.}\ }\textbf {\bibinfo
  {volume} {122}},\ \bibinfo {pages} {040504} (\bibinfo {year}
  {2019})}\BibitemShut {NoStop}%
\bibitem [{\citenamefont {Hubregtsen}\ \emph {et~al.}(2022)\citenamefont
  {Hubregtsen}, \citenamefont {Wierichs}, \citenamefont {Gil-Fuster},
  \citenamefont {Derks}, \citenamefont {Faehrmann},\ and\ \citenamefont
  {Meyer}}]{hubregtsen2022training}%
  \BibitemOpen
  \bibfield  {author} {\bibinfo {author} {\bibfnamefont {T.}~\bibnamefont
  {Hubregtsen}}, \bibinfo {author} {\bibfnamefont {D.}~\bibnamefont
  {Wierichs}}, \bibinfo {author} {\bibfnamefont {E.}~\bibnamefont
  {Gil-Fuster}}, \bibinfo {author} {\bibfnamefont {P.-J.~H.}\ \bibnamefont
  {Derks}}, \bibinfo {author} {\bibfnamefont {P.~K.}\ \bibnamefont
  {Faehrmann}},\ and\ \bibinfo {author} {\bibfnamefont {J.~J.}\ \bibnamefont
  {Meyer}},\ }\bibfield  {title} {\bibinfo {title} {Training quantum embedding
  kernels on near-term quantum computers},\ }\href
  {https://doi.org/https://doi.org/10.1103/PhysRevA.106.042431} {\bibfield
  {journal} {\bibinfo  {journal} {Phys. Rev. A}\ }\textbf {\bibinfo {volume}
  {106}},\ \bibinfo {pages} {042431} (\bibinfo {year} {2022})}\BibitemShut
  {NoStop}%
\bibitem [{\citenamefont {Bowie}\ \emph {et~al.}(2023)\citenamefont {Bowie},
  \citenamefont {Shrapnel},\ and\ \citenamefont {Kewming}}]{bowie2023quantum}%
  \BibitemOpen
  \bibfield  {author} {\bibinfo {author} {\bibfnamefont {C.}~\bibnamefont
  {Bowie}}, \bibinfo {author} {\bibfnamefont {S.}~\bibnamefont {Shrapnel}},\
  and\ \bibinfo {author} {\bibfnamefont {M.~J.}\ \bibnamefont {Kewming}},\
  }\bibfield  {title} {\bibinfo {title} {Quantum kernel evaluation via
  {H}ong–{O}u–{M}andel interference},\ }\href
  {https://doi.org/https://doi.org/10.1088/2058-9565/acfba9} {\bibfield
  {journal} {\bibinfo  {journal} {Quantum Sci. Technol.}\ }\textbf {\bibinfo
  {volume} {9}},\ \bibinfo {pages} {015001} (\bibinfo {year}
  {2023})}\BibitemShut {NoStop}%
\bibitem [{\citenamefont {Sergioli}\ \emph {et~al.}(2019)\citenamefont
  {Sergioli}, \citenamefont {Giuntini},\ and\ \citenamefont
  {Freytes}}]{sergioli2019new}%
  \BibitemOpen
  \bibfield  {author} {\bibinfo {author} {\bibfnamefont {G.}~\bibnamefont
  {Sergioli}}, \bibinfo {author} {\bibfnamefont {R.}~\bibnamefont {Giuntini}},\
  and\ \bibinfo {author} {\bibfnamefont {H.}~\bibnamefont {Freytes}},\
  }\bibfield  {title} {\bibinfo {title} {A new quantum approach to binary
  classification},\ }\href
  {https://doi.org/https://doi.org/10.1371/journal.pone.0216224} {\bibfield
  {journal} {\bibinfo  {journal} {PloS One}\ }\textbf {\bibinfo {volume}
  {14}},\ \bibinfo {pages} {e0216224} (\bibinfo {year} {2019})}\BibitemShut
  {NoStop}%
\bibitem [{\citenamefont {Blank}\ \emph {et~al.}(2020)\citenamefont {Blank},
  \citenamefont {Park}, \citenamefont {Rhee},\ and\ \citenamefont
  {Petruccione}}]{blank2020quantum}%
  \BibitemOpen
  \bibfield  {author} {\bibinfo {author} {\bibfnamefont {C.}~\bibnamefont
  {Blank}}, \bibinfo {author} {\bibfnamefont {D.~K.}\ \bibnamefont {Park}},
  \bibinfo {author} {\bibfnamefont {J.-K.~K.}\ \bibnamefont {Rhee}},\ and\
  \bibinfo {author} {\bibfnamefont {F.}~\bibnamefont {Petruccione}},\
  }\bibfield  {title} {\bibinfo {title} {Quantum classifier with tailored
  quantum kernel},\ }\href
  {https://doi.org/https://doi.org/10.1038/s41534-020-0272-6} {\bibfield
  {journal} {\bibinfo  {journal} {npj Quantum Inf.}\ }\textbf {\bibinfo
  {volume} {6}},\ \bibinfo {pages} {41} (\bibinfo {year} {2020})}\BibitemShut
  {NoStop}%
\bibitem [{\citenamefont {Nagies}\ \emph {et~al.}(2026)\citenamefont {Nagies},
  \citenamefont {Tolotti}, \citenamefont {Pastorello},\ and\ \citenamefont
  {Blanzieri}}]{nagies2026enhancing}%
  \BibitemOpen
  \bibfield  {author} {\bibinfo {author} {\bibfnamefont {S.}~\bibnamefont
  {Nagies}}, \bibinfo {author} {\bibfnamefont {E.}~\bibnamefont {Tolotti}},
  \bibinfo {author} {\bibfnamefont {D.}~\bibnamefont {Pastorello}},\ and\
  \bibinfo {author} {\bibfnamefont {E.}~\bibnamefont {Blanzieri}},\ }\href@noop
  {} {\bibinfo {title} {Enhancing expressivity of quantum neural networks based
  on the swap test}} (\bibinfo {year} {2026}),\ \Eprint
  {https://arxiv.org/abs/2506.16938} {arXiv:2506.16938 [quant-ph]} \BibitemShut
  {NoStop}%
\bibitem [{\citenamefont {Bishop}\ and\ \citenamefont
  {Nasrabadi}(2006)}]{book:Bishop}%
  \BibitemOpen
  \bibfield  {author} {\bibinfo {author} {\bibfnamefont {C.~M.}\ \bibnamefont
  {Bishop}}\ and\ \bibinfo {author} {\bibfnamefont {N.~M.}\ \bibnamefont
  {Nasrabadi}},\ }\href@noop {} {\emph {\bibinfo {title} {Pattern Recognition
  and Machine Learning}}},\ Vol.~\bibinfo {volume} {4}\ (\bibinfo  {publisher}
  {Springer},\ \bibinfo {year} {2006})\BibitemShut {NoStop}%
\bibitem [{\citenamefont {Roncallo}\ \emph
  {et~al.}(2025{\natexlab{a}})\citenamefont {Roncallo}, \citenamefont
  {Morgillo}, \citenamefont {Macchiavello}, \citenamefont {Maccone},\ and\
  \citenamefont {Lloyd}}]{roncallo2025quantum}%
  \BibitemOpen
  \bibfield  {author} {\bibinfo {author} {\bibfnamefont {S.}~\bibnamefont
  {Roncallo}}, \bibinfo {author} {\bibfnamefont {A.~R.}\ \bibnamefont
  {Morgillo}}, \bibinfo {author} {\bibfnamefont {C.}~\bibnamefont
  {Macchiavello}}, \bibinfo {author} {\bibfnamefont {L.}~\bibnamefont
  {Maccone}},\ and\ \bibinfo {author} {\bibfnamefont {S.}~\bibnamefont
  {Lloyd}},\ }\bibfield  {title} {\bibinfo {title} {Quantum optical classifier
  with superexponential speedup},\ }\href
  {https://doi.org/https://doi.org/10.1038/s42005-025-02020-5} {\bibfield
  {journal} {\bibinfo  {journal} {Commun. Phys}\ }\textbf {\bibinfo {volume}
  {8}},\ \bibinfo {pages} {147} (\bibinfo {year}
  {2025}{\natexlab{a}})}\BibitemShut {NoStop}%
\bibitem [{\citenamefont {Roncallo}\ \emph
  {et~al.}(2025{\natexlab{b}})\citenamefont {Roncallo}, \citenamefont
  {Morgillo}, \citenamefont {Lloyd}, \citenamefont {Macchiavello},\ and\
  \citenamefont {Maccone}}]{roncallo2025shallow}%
  \BibitemOpen
  \bibfield  {author} {\bibinfo {author} {\bibfnamefont {S.}~\bibnamefont
  {Roncallo}}, \bibinfo {author} {\bibfnamefont {A.~R.}\ \bibnamefont
  {Morgillo}}, \bibinfo {author} {\bibfnamefont {S.}~\bibnamefont {Lloyd}},
  \bibinfo {author} {\bibfnamefont {C.}~\bibnamefont {Macchiavello}},\ and\
  \bibinfo {author} {\bibfnamefont {L.}~\bibnamefont {Maccone}},\ }\href@noop
  {} {\bibinfo {title} {Quantum optical shallow networks}} (\bibinfo {year}
  {2025}{\natexlab{b}}),\ \Eprint {https://arxiv.org/abs/2507.21036}
  {arXiv:2507.21036} \BibitemShut {NoStop}%
\bibitem [{\citenamefont {Roncallo}\ and\ \citenamefont
  {Morgillo}()}]{rep:QON}%
  \BibitemOpen
  \bibfield  {author} {\bibinfo {author} {\bibfnamefont {S.}~\bibnamefont
  {Roncallo}}\ and\ \bibinfo {author} {\bibfnamefont {A.~R.}\ \bibnamefont
  {Morgillo}},\ }\href@noop {} {}\bibinfo {note}
  {\url{https://github.com/simoneroncallo/hybrid-multinomial-classifier}}\BibitemShut
  {NoStop}%
\end{thebibliography}%
\end{document}